\pdfoutput=1
\documentclass[twocolumn,superscriptaddress,aps,preprintnumbers,amsmath,amssymb,prl,nofootinbib]{revtex4-2}

\usepackage{amsmath}
\usepackage{amssymb}
\usepackage{amsfonts}
\usepackage{graphicx}
\usepackage{xcolor}
\usepackage{xfrac}
\usepackage{comment}
\usepackage{pifont}
\usepackage{physics}
\usepackage{fourier}
\usepackage{hyperref}
\usepackage{bm}
\usepackage{enumitem}

\definecolor{rossoferrari}{HTML}{D9073D}
\definecolor{mediumblue}{HTML}{0000CD}
\definecolor{forestgreen}{HTML}{228B22}
\definecolor{desy_blue}{HTML}{009EE2}
\definecolor{desy_orange}{HTML}{FD8800}
\definecolor{light_pink}{rgb}{1,0.4,0.4}
\definecolor{light_blue}{rgb}{0.284602,0.317763,0.963947}
\hypersetup{setpagesize=false,bookmarksnumbered=true,bookmarksopen=true,%
colorlinks=true,linkcolor=light_blue,urlcolor=rossoferrari,citecolor=rossoferrari,linktocpage=false}

\newcommand{\mat}[1]{\bm{\underline{#1}}}

\newcommand{\MeV}{\,\mathrm{MeV}}
\newcommand{\GeV}{\,\mathrm{GeV}}

\begin{document}


\title{
Baryon asymmetry of the Universe from lepton flavor violation
}

\author{Kyohei Mukaida}
\email{kyohei.mukaida@kek.jp}
\affiliation{Theory Center, IPNS, KEK, Tsukuba, Ibaraki 305-0801, Japan}
\affiliation{Graduate University for Advanced Studies, Tsukuba, Ibaraki 305-0801, Japan}

\author{Kai Schmitz}
\email{kai.schmitz@cern.ch}
\affiliation{Theoretical Physics Department, CERN, 1211 Geneva 23, Switzerland}

\author{Masaki Yamada}
\email{m.yamada@tohoku.ac.jp}
\affiliation{FRIS, Tohoku University, Sendai, Miyagi 980-8578, Japan}
\affiliation{Department of Physics, Tohoku University, Sendai, Miyagi 980-8578, Japan}

\preprint{KEK-TH-2364, CERN-TH-2021-183, TU-1137}

\date{\today}


\begin{abstract}
\noindent
Charged-lepton flavor violation (CLFV) is a smoking-gun signature of physics beyond the Standard Model. The discovery of CLFV in upcoming experiments would indicate that CLFV processes must have been efficient in the early Universe at relatively low temperatures. In this letter, we point out that such efficient CLFV interactions open up new ways of creating the baryon asymmetry of the Universe. First, we quote the two-loop corrections from charged-lepton Yukawa interactions to the chemical transport in the Standard Model plasma, which imply that nonzero lepton flavor asymmetries summing up to $B-L = 0$ are enough to generate the baryon asymmetry. Then, we describe two scenarios of what we call {\it leptoflavorgenesis}, where efficient CLFV processes are responsible for the generation of primordial lepton flavor asymmetries that are subsequently converted to a baryon asymmetry by weak sphaleron processes. Here, the conversion factor from lepton flavor asymmetry to baryon asymmetry is suppressed by charged-lepton Yukawa couplings squared, which provides a natural explanation for the smallness of the observed baryon-to-photon ratio.
\end{abstract}


\maketitle


\textit{\textbf{Introduction.\,---\,}} The Standard Model (SM) of particle physics has been established observationally after the discovery of Higgs boson. Its \textit{classical} action enjoys the accidental global symmetry $U(1)_{B+L} \times U(1)_{L_e - L_\mu} \times U(1)_{L_\mu - L_\tau} \times  U (1)_{B-L}$, corresponding to the conservation of the baryon charge $B$ and the flavored lepton charges $L_{e, \mu, \tau}$. \textit{Quantum mechanically}, $U(1)_{B+L}$ is violated by the chiral anomaly~\cite{tHooft:1976rip}. Although suppressed in the vacuum, $B+L$ violation by means of weak sphaleron processes is efficient at temperatures $10^2\,\mathrm{GeV}  \lesssim T\lesssim 10^{12}\,\mathrm{GeV}$~\cite{Dimopoulos:1978kv,Manton:1983nd,Klinkhamer:1984di, Kuzmin:1985mm}.
Also, the discovery of neutrino oscillations revealed that the lepton flavor symmetries, $U(1)_{L_e - L_\mu} \times U(1)_{L_\mu - L_\tau}$, are violated~\cite{Pontecorvo:1967fh,Gribov:1968kq,Super-Kamiokande:1998kpq,SNO:2001kpb,SNO:2002tuh}, which further implies the violation of $U(1)_{B-L}$ if neutrinos are Majorana fermions~\cite{Weinberg:1979sa}. However, such interactions are so feeble that they could be inefficient up to $T_{B-L} \sim 10^{13}\, \mathrm{GeV}$, depending on possible UV completions of the neutrino sector.

Charged-lepton flavor violation (CLFV) is currently attracting a lot of attention~\cite{Kuno:1999jp,Calibbi:2017uvl,Babu:2020hun,Aloni:2021wzk,Calibbi:2021pyh, Bjorkeroth:2018dzu,
Bauer:2019gfk,Cornella:2019uxs,Endo:2020mev,Ishida:2020oxl,Calibbi:2020jvd,Escribano:2020wua,
DiLuzio:2020oah,Bonnefoy:2020llz,Ma:2021jkp,Bauer:2021mvw}, 
since its discovery would undoubtedly imply physics beyond the SM, 
enabling us to probe new physics at extremely high energy scales, such as $\order{10^8}\, \mathrm{GeV}$ by $\mu \to e \gamma$~\cite{Calibbi:2017uvl} and $\order{10^9}\, \mathrm{GeV}$ by $\mu \to e a$, where $a$ is a pseudo-Nambu--Goldstone boson like an axion or familon~\cite{Feng:1997tn,Calibbi:2020jvd}.%
\footnote{Anomalies and hints of lepton flavor universality violation in observables such as rare $B$ meson decays~\cite{LHCb:2017avl,LHCb:2019hip,LHCb:2021trn} (see also Refs.~\cite{HFLAV:2019otj,Altmannshofer:2020axr}) and the anomalous magnetic moment of the muon $(g-2)_\mu$~\cite{Muong-2:2006rrc,Jegerlehner:2009ry,Muong-2:2021ojo} may also be linked to new CLFV interactions~\cite{Feng:2001sq,Calibbi:2006nq,Giudice:2012ms,Glashow:2014iga} (see, \textit{e.g.}, Refs.~\cite{Greljo:2021xmg,Calibbi:2021qto,Ibe:2021cvf,Li:2021lnz,Buras:2021btx,BhupalDev:2021ipu,Greljo:2021npi,Hou:2021qmf}).}
Furthermore, if CLFV should be observed in upcoming experiments, we would learn that CLFV interactions must have been efficient at relatively low temperatures in the early Universe. For instance, the discovery of $\mu \to e \gamma$ would imply that $U(1)_{L_e - L_\mu}$ violating interactions are efficient at $T \gtrsim 10^4 \, \mathrm{GeV}$~\cite{Deppisch:2015yqa}. Above this temperature, LFV interactions then enforce nontrivial relations among the lepton chemical potentials, \textit{e.g.}, 
$\mu_{\mu} - \mu_{e} = 0$, 
leading to a different chemical equilibrium at that temperature. This is in contrast to the discovery of neutrino oscillations, where $T_{B-L}$ is not necessarily low but could rather be as high as $10^{13}\, \mathrm{GeV}$.

In this letter, we point out that the discovery of CLFV processes would change the paradigm of baryogenesis. As history shows, the discovery of the violation of SM global symmetries has repeatedly opened up new baryogenesis mechanisms. For instance, the weak sphaleron dramatically changes the chemical transport before and after the electroweak phase transition (EWPT), leading to two famous scenarios; baryogenesis at the EWPT~\cite{Kuzmin:1985mm} and baryogenesis through leptogenesis ($B-L$ genesis) before the EWPT~\cite{Fukugita:1986hr}. In a similar spirit, we will now present conceptually new possibilities for baryogenesis.

We first note that the charged-lepton Yukawa couplings are hierarchical. This implies that the rates for the conversion of the flavored $B/3-L_f$ charges into the baryon charge at sphaleron decoupling are not universal for different flavors. We thus do not necessarily need a $B-L$ asymmetry; a lepton flavor asymmetry alone is enough to obtain a nonzero baryon asymmetry during the EWPT, though the conversion factor is suppressed by the charged-lepton Yukawa coupling squared~\cite{Kuzmin:1987wn,Khlebnikov:1988sr,Laine:1999wv}. Therefore, we only need to generate some lepton flavor asymmetry at a temperature higher than the electroweak scale\,---\,a process that we will refer to as \textit{leptoflavorgenesis} in the following. 

We emphasize that leptoflavorgenesis does not represent yet another variant of leptogenesis. In particular, it does not refer to flavor effects in leptogenesis~\cite{Pilaftsis:2005rv,Abada:2006fw,Nardi:2006fx,Abada:2006ea,Dev:2017trv} (see also Refs.~\cite{Pilaftsis:2004xx,Deppisch:2015yqa} for the effect of CLFV in leptogenesis), where lepton flavor asymmetry plays an important role due to the flavor-dependent wash-out effects of right-handed neutrinos in the presence of a nonvanishing total $B-L$ asymmetry. Leptoflavorgenesis is not related to right-handed neutrinos nor does it generate a $B-L$ asymmetry. 
Note also that leptoflavorgenesis generate baryon asymmetry from $B-L=0$ in the SM (and whole) sector, which is different from the Dirac leptogenesis~\cite{Dick:1999je} and cloistered baryogenesis~\cite{AristizabalSierra:2013lyx}, where they generate $(B-L)$ in the SM sector with the opposite $(B-L)$ in a sector that is decoupled from the electroweak sphaleron.

Leptoflavorgenesis is roughly classified into two cases, \textit{i.e.}, (i) lepton flavor asymmetry generation at the decoupling of LFV interactions, and (ii) lepton flavor asymmetry generation through other asymmetries before the decoupling of LFV interactions, analogous to the situation for electroweak baryogenesis and leptogenesis. Unfortunately, in the first option (i), it is difficult to directly generate a lepton flavor asymmetry by using LFV interactions that would be discovered in the near future. (See Ref.~\cite{March-Russell:1999hpw} as an example of this scenario.) This is because we need a source of large CP violation in order to generate enough asymmetry while the discovery of LFV does not necessarily provide such a CP-violating source. We shall therefore concentrate on the latter scenario (ii) throughout this letter.

We propose two concrete leptoflavorgenesis scenarios operative before the decoupling of LFV interactions; wash-in leptoflavorgenesis and spontaneous leptoflavorgenesis. 
We first need to generate some asymmetries that are then converted to the lepton flavor asymmetry through LFV interactions. As emphasized in~\cite{Domcke:2020kcp}, there exist approximately conserved charges in the SM for $T \gtrsim 10^5 \, \mathrm{GeV}$, as the SM interactions become less effective than the cosmic expansion at higher temperatures. Therefore, the production of such charges via UV physics suffices, which we refer to as wash-in leptoflavorgenesis in analogy to wash-in leptogenesis~\cite{Domcke:2020quw} 
(see also Refs.~\cite{Antaramian:1993nt, Fong:2015vna}). Another example of generating some asymmetries is to couple an axion-like particle (ALP) to a SM current that is not conserved (see, \textit{e.g.}, Ref.~\cite{Domcke:2020kcp}). If the ALP has nonzero velocity, it acts as an effective chemical potential for the SM plasma, generically leading to  lepton flavor asymmetries. We refer to this mechanism as spontaneous leptoflavorgenesis.

\smallskip
\textit{\textbf{Decoupling temperature of LFV.}}\,---\, Let us first estimate the decoupling temperatures of some LFV interactions. 
We are interested in the processes that are accessible by collider experiments or astrophysical observations in the near future, such as the one like $\ell_f \to \ell_{f'} X$, where $\ell_f$ is the $f$-th generation charged lepton and $X$ represents a neutral particle.

For example, $\mu \to e \gamma$ is induced by the following dimension-six operator 
(see, \textit{e.g.}, Refs.~\cite{Calibbi:2017uvl})
\begin{equation}
 \frac{2 C^{f f'}_{\ell W}}{\Lambda^2} L^\dag_{Lf} \sigma^{\mu\nu} e_{Rf'} W_{\mu\nu} \Phi
 + \frac{C^{f f'}_{\ell B}}{\Lambda^2}  L_{Lf}^\dag \sigma^{\mu\nu} e_{Rf'}  B_{\mu\nu} \Phi + \text{H.c.},
 \label{eq:egam}
\end{equation}
where $C^{f f'}_{\ell W}$ and $C^{f f'}_{\ell B}$ are dimensionless coefficients,
$\Lambda$ is the cutoff scale of the operator, 
$L_f$ is the left-handed lepton doublet of the $f$-th flavor,
$e_f$ is the right-handed lepton singlet of the $f$-th flavor,
$W_{\mu\nu}$ ($B_{\mu\nu}$) is the field strength of $SU(2)_W$ ($U(1)_Y$),
and $\Phi$ is the Higgs doublet.
After electroweak symmetry breaking, this operator induces
\begin{equation}
  \frac{C^{f f'}_{\ell \gamma}}{\Lambda^2} \frac{v}{\sqrt{2}} 
  \overline{\ell}_f \sigma^{\mu\nu} P_R \ell_{f'} F_{\mu\nu} + \text{H.c.},
\end{equation}
where we define $C^{f f'}_{\ell\gamma} \equiv \cos \theta_W C^{f f'}_{\ell  B} - \sin \theta_W C^{f f'}_{\ell W}$ with $\theta_W$ being the Weinberg angle and the field strength of $U(1)_\text{EM}$ is denoted by $F_{\mu\nu}$. One may easily see that Eq.~\eqref{eq:egam} does not violate $U(1)_{B-L}$ but does violate $U(1)_{L_f - L_{f'}}$. This operator is constrained by current experiments; the tightest bound comes from the nonobservation of $\mu \to e \gamma$: 
${\rm Br}(\mu^+ \to e^+ \gamma) \le 4.2 \times 10^{-13}$~\cite{MEG:2016leq}. 
This leads to $\Lambda / \left( (C_{\ell \gamma}^{\mu e})^2 + (C_{\ell \gamma}^{e \mu})^2 \right)^{1/4} \gtrsim 6.7 \times 10^7 \,\mathrm{GeV}$~\cite{Calibbi:2017uvl}.

The LFV interaction rate above the electroweak scale is
\begin{equation}
  \gamma_{\ell W/B}^{f f'} \simeq 
  \frac{336 T^5}{\pi^5 \Lambda^4} 
  \left( 3 \abs{C_{\ell W}^{f f'}}^2 + \abs{C_{\ell B}^{f f'}}^2 \right). 
\end{equation}
Since the temperature dependence is stronger than that of the Hubble rate, $H \propto T^2$, the LFV interaction is in equilibrium at early times before it decouples at low temperatures. The decoupling temperature of $L_f - L_{f'}$ violation is defined by the temperature at $2 \gamma_{\ell\gamma}^{f f'} = 4/7 H$, where the factor of $2$ comes from $f \leftrightarrow f'$, and the factor of $4/7$ comes from the convention used in Ref.~\cite{Domcke:2020kcp}. This is then estimated as%
\footnote{
If this is higher than the mass of integrated heavy field for the effective operator, 
the decoupling temperature should be calculated in the UV theory. The transport equation we will use should also be written in the UV model.
}
\begin{equation}
  T^{\rm dec}_{\ell \gamma} \simeq 
  3.1 \times 10^4 \,\mathrm{GeV} \, 
  \qty( \frac{\Lambda / \sqrt{C_{\ell \gamma}}}{10^8 \, \mathrm{GeV}} )^{4/3}.
\end{equation}
where we assume $C_{\ell W}^{f f'} = 0$ with a universal coupling for $C_{\ell B}^{f f'} \simeq 1.1 C_{\ell \gamma}$. 
In the near future, the MEG II experiment will reach a sensitivity of ${\rm Br}(\mu^+ \to e^+ \gamma) = 6 \times 10^{-14}$, which is going to probe $\Lambda / C_{e\gamma}^{1/2}$ up to $1.0 \times 10^8 \GeV$~\cite{MEGII:2018kmf,MEGII:2021fah} (see also Ref.~\cite{Cavoto:2017kub}). Therefore, if $\mu \to e \gamma$ should be discovered in upcoming experiments, we will learn that LFV interactions are equilibrated at temperatures above $\mathcal{O}(10^4) \,\mathrm{GeV}$.

One may also consider a LFV interaction of the type $l_f \to l_{f'} a$ via an operator of the form
\begin{equation}
\label{eq:muea}
 \frac{\partial_\mu a}{2 f_a}\,
 \qty( C^{f f'}_{La} L^\dag_{Lf} \sigma^\mu L_{L f'}
 +
 C^{f f'}_{Ra} e^\dag_{Rf} \overline{\sigma}^\mu e_{R f'}
 ),
\end{equation}
where $a$ is a pseudo-Nambu--Goldstone boson like an axion and familon, $f_a$ is its decay constant, and $C^{f f'}_{La}$ and $C^{f f'}_{Ra}$ are dimensionless coefficients (see, \textit{e.g.}, Refs.~\cite{Calibbi:2020jvd,Bauer:2021mvw}).
The current and expected future bounds on the effective axion decay constant, 
$F_{V/Aa}^{f f'} \equiv 2 f_a/\sqrt{|C^{ff'}_{Va}|^2+|C^{ff'}_{Aa}|^2}$ 
with $C^{ff'}_{V/A} \equiv ( C^{f f'}_{Ra} \mp C^{f f'}_{La})/2$, 
are about $4.8 \times 10^{9} \, \mathrm{GeV}$ by Jodidio \textit{et al.}~\cite{Jodidio:1986mz,Feng:1997tn} and $2.9 \times 10^{10} \GeV$ by MEGII-fwd~\cite{Calibbi:2020jvd} and Mu3e-online~\cite{Perrevoort:2018okj}, respectively, though the precise value depends on the chirality of the interactions. 
The corresponding scattering rate of flavor-changing process is given by 
\begin{align}
  &\gamma_{\ell_L a}^{f f'} \simeq 
  \frac{24 T^3}{\pi^4 f_a^2} 
  \left( \alpha_Y + 6 \alpha_2 \right) 
  \abs{C_{L a}^{ff'}}^2
  \\
  &\gamma_{\ell_R a}^{f f'} \simeq 
  \frac{24  T^3}{\pi^4 f_a^2} 
  \left( 4\alpha_Y \right)  \abs{C_{R a}^{ff'}}^2, 
\end{align}
for the left- and right-handed leptons, respectively. 
Hence, if $\mu \to e a$ is discovered in near-future measurements, the corresponding LFV interaction is equilibrated above
\begin{equation}
 T^{\rm dec}_{\ell_L a} \simeq 7.2 \times 10^2 \, \mathrm{GeV}\, \qty( \frac{F_a}{10^{10} \GeV} )^2
\end{equation}
where we take a universal coupling $C^{ff'}_{La} = 1$, $C^{ff'}_{Ra} = 0$ as an example and where the decoupling temperature in this case is defined by the temperature at $2 \gamma_{\ell_L a}^{f f'} = H$.

In both examples, the decoupling temperature is higher than the electroweak scale. Therefore, the three flavored $B-L$ charges, \textit{i.e.}, $\Delta_f \equiv B/3 - L_f$, become conserved by the time of the electroweak crossover. 


\smallskip
\textit{\textbf{Baryon charge transport during the EWPT.}}\,---\,
In the SM, the EWPT proceeds as a crossover~\cite{Csikor:1998eu}, 
where the weak sphaleron process decouples at $T_\text{Sp} \approx 135 \GeV$~\cite{DOnofrio:2014rug,Kamada:2016cnb}
with the neutral component of Higgs field value 
being 
$x_\text{Sp} \equiv \left. \left<\phi \right> / T \right|_\text{Sp} \approx 1.2$~\cite{DOnofrio:2015gop,Kamada:2016eeb}.

During the EWPT, 
we have in total three conserved charges: $\Delta_e = B/3 - L_e$, $\Delta_\mu = B/3 - L_\mu$, and $\Delta_\tau = B/3 - L_\tau$.  In the literature, it is widely studied how $B-L \neq 0$ is converted to $B\neq 0$ by weak sphalerons. Here, we consider another possibility, assuming that there is nonzero $\Delta_f$ ($f = e, \mu, \tau$) but zero $B-L = \sum_f \Delta_f$. In fact, as shown in Refs.~\cite{Kuzmin:1987wn,Khlebnikov:1988sr,Laine:1999wv}, the hierarchies in the charged-lepton Yukawa couplings lead to different conversion factors for each flavor. As a result, the baryon asymmetry after sphaleron decoupling is given by
\begin{equation}
 Y_B \simeq \frac{3 A \qty( x_\text{Sp} )}{13 \pi^2} \sum_{f=e,\mu,\tau} y_{f}^2 Y_{\Delta_f}, 
 \label{eq:nb1}
\end{equation}
for vanishing total $B-L$, where
\begin{equation}
A \qty(x) \equiv \frac{13 (1034 + 2473 x^2 + 792 x^4)}{48 (869 + 333 x^2)} \,,\quad A(x_\text{Sp}) \simeq 1.3 \,.
\end{equation}
We thus only need to generate some lepton flavor asymmetry before the EWPT. The conversion factor is suppressed by the charged-lepton Yukawa coupling: $3 A \qty( x_\text{Sp} ) y_\mu^2 / (13 \pi^2) \simeq 1.1 \times 10^{-8}$ for $\Delta_\mu$ and $3 A \qty( x_\text{Sp} ) y_\tau^2 / (13 \pi^2) \simeq 3.0 \times 10^{-6}$ for $\Delta_\tau$. As the observed baryon asymmetry is 
$Y_B \simeq 9 \times 10^{-11}$, successful leptoflavorgenesis requires $Y_{\Delta_\mu} \simeq 8 \times 10^{-3}$ or $Y_{\Delta_\tau} \simeq 3 \times 10^{-5}$.


\smallskip
\textit{\textbf{Transport equations with LFV interactions.}}\,---\,We can greatly simplify the transport equations around the LFV decoupling temperature, $T^{\rm dec}_{\ell \gamma/a} \sim 10^{4\cdots5} \GeV$, assuming that all other interactions except for the electron Yukawa and LFV interactions are equilibrated. The transport equations for the right-handed electron charge density $q_e$ and $\Delta_f$ are then given by 
\begin{align}
 \dot{q}_e + 3 H q_e &= - \frac{1}{T} \gamma_e 
\left( \mu_e - \mu_{\ell_e} + \mu_\phi + \mu_{\rm bias}^e
\right) 
\notag\\
&\quad 
- \sum_{f'} \frac{1}{T} \gamma_{\ell W/B}^{f'e} 
\left( \mu_{e} - \mu_{\ell_{f'}} + \mu_\phi + \mu_{\rm bias}^{\ell W/B, f'e} 
\right)
\\
\dot{q}_{\Delta_f} + 3 H q_{\Delta_f} &= - \sum_{f'} \frac{1}{T} \gamma_{\ell W/B}^{ff'} 
\left( \mu_{f'} - \mu_{\ell_{f}} + \mu_\phi + \mu_{\rm bias}^{\ell W/B, ff'} 
\right)
\notag\\
&\quad - \sum_{f'} \frac{1}{T} \gamma_{\ell W/B}^{f' f} 
\left( - \mu_{f} + \mu_{\ell_{f'}} - \mu_\phi - \mu_{\rm bias}^{\ell W/B, f'f} 
\right), 
\end{align}
for the LFV interaction in Eq.~(\ref{eq:egam}), where $\mu_i$ is the chemical potential of species $i$. The bias factors $\mu_{\rm bias}^I$ are only relevant for spontaneous leptoflavorgenesis (see below), where the index $I$ represents interactions. The electron Yukawa interaction rate $\gamma_e$ is given by $\gamma_e/H \simeq 4/7 T_{y_e}^{\rm dec}/T$, where $T_{y_e}^{\rm dec} \simeq 1.1 \times 10^5 \GeV$ is its decoupling temperature~\cite{Garbrecht:2014kda,Bodeker:2019ajh,Domcke:2020kcp}. We can rewrite $\mu_{\ell_f}$, $\mu_{f'}$, and $\mu_\phi$ in terms of $\mu_e$ and $\mu_{\Delta_f}$ by integrating out the spectator processes, as explained in the Supplemental Material (see also Refs.~\cite{Domcke:2020kcp,Domcke:2020quw}). 
The resulting transport equations are symmetric under $\mu \leftrightarrow \tau$. Together with $B-L$ conservation, we obtain $q_{\Delta_\mu} = q_{\Delta_\tau} = -(1/2)q_{\Delta_e}$, if the bias factors (see below) and initial charges are also symmetric. 
The relevant transport equations are therefore reduced to 
\begin{align}
\label{eq:transport1-2}
 \dot{q}_e + 3 H q_e &= - \frac{1}{T} \gamma_e 
\left( \frac{711}{481} \mu_e - \frac{474}{481} \mu_{\Delta_\tau} - \sum_I b_e^I \mu_{\rm bias}^I 
\right), 
\\
\label{eq:transport2-2}
\dot{q}_{\Delta_\tau} + 3 H q_{\Delta_\tau} &= - \frac{1}{T} \gamma_{\ell B} 
\left( 
 - \frac{237}{481} \mu_e 
+ \frac{639}{481} \mu_{\Delta_\tau} - \sum_I b_{\Delta_\tau}^I \mu_{\rm bias}^I 
\right), 
\end{align}
where we assume $C_{\ell W}^{f f'} = 0$ and a universal coupling $C_{\ell B}^{f f'} = C_{\ell B} \simeq 1.1 C_{\ell \gamma}$ except for $f'=e$ for simplicity. 
Here we assume that the right-handed electric charge is not washed out by the CLFV process by taking $C_{\ell B}^{f e} = 0$, which is required for wash-in leptoflavorgenesis to work. 
The explicit values of the coefficients $b_e^I$ and $b_{\Delta_\tau}^I$ for the bias factors are given in the Supplemental Material.
For the case of the LFV interaction in Eq.~(\ref{eq:muea}), the transport equations are given by the same form with the replacement of $\gamma_{\ell B} \to \gamma_{\ell_L a}$ 
for $C^{ff'}_{La} = 1$, $C^{ff'}_{Ra} = 0$. 


\smallskip
\textit{\textbf{Wash-in leptoflavorgenesis.}}\,---\,Now we shall study the generation of flavored $B-L$ charges $\Delta_f$ via the transport equations. Let us first consider the case without bias factors ($\mu_{\rm bias}^I =0$), while the initial right-handed electron charge is nonzero ($\mu^{\rm ini}_C \ne 0$ for $C = e$). 
In this case, the electron charge is converted to $q_{\Delta_f}$ as described by the above transport equations. The resulting $q_{\Delta_f}$ is conserved at $T \lesssim T_{\ell \gamma/a}^{\rm dec}$. Taking into account Eq.~(\ref{eq:nb1}), we obtain a nonzero baryon charge. We dub this scenario wash-in leptoflavorgenesis. The initial electron charge is expected to be generated by some other mechanism, such as the Affleck--Dine mechanism with a $B-L = 0$ (\textit{e.g.}, $u_Ru_Rd_Re_R$) flat direction ~\cite{Affleck:1984fy,Dine:1995uk,Dine:1995kz}. Another possibility is axion-inflation with a strong Chern--Simons coupling~\cite{Adshead:2015kza,Adshead:2018oaa,Domcke:2018eki,Domcke:2019mnd}. We note that preexisting flavored $B-L$ charges, if any, are washed out by the LFV interactions at $T \gtrsim T_{\ell \gamma/a}^{\rm dec}$, and the results below are independent of the initial $\Delta_f$ charges. 

If the LFV interaction is strong enough, the parenthesis on the right-hand side of Eq.~(\ref{eq:transport2-2}) is forced to vanish. Here, note that the Yukawa (LFV) interactions enter (leave) equilibrium as the temperature drops. If the LFV process is decoupled while the electron Yukawa interaction is negligible, we obtain the result in the so-called strong wash-in regime,
$\mu_{\Delta_\tau} = (79/213) \mu_e^{\rm ini}$.

\begin{figure}[t]
	\centering
 	\includegraphics[width=0.9\linewidth]{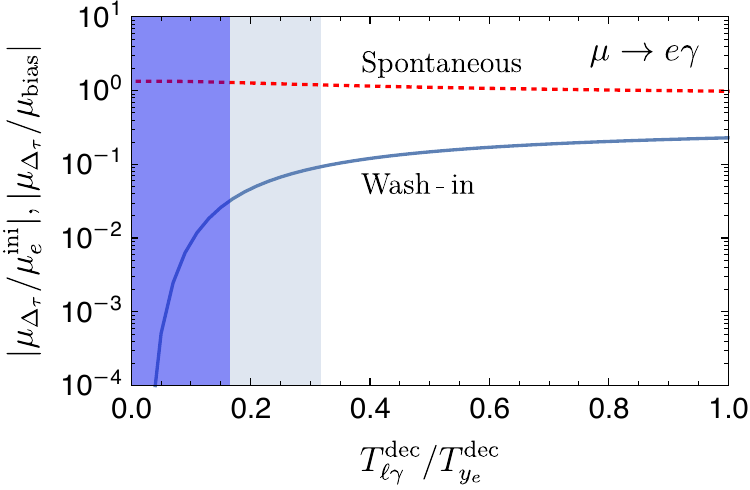} 
 	\vspace{0.6cm}\\
 	\includegraphics[width=0.9\linewidth]{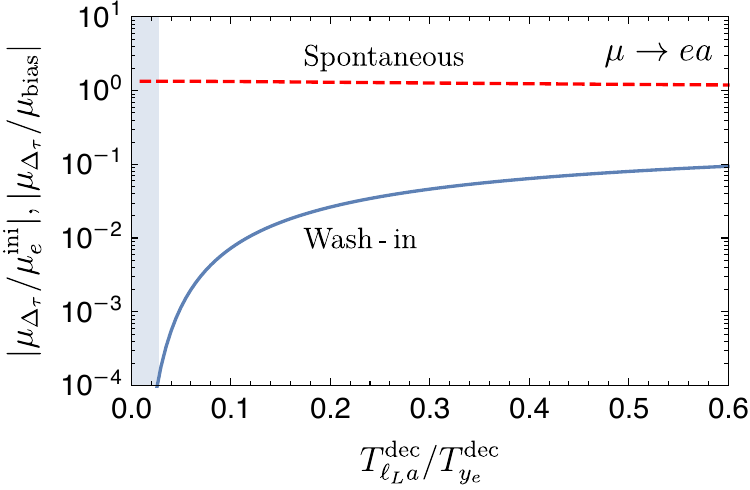} 
	\caption{ 
	$|\mu_{\Delta_\tau}/\mu_e^{\rm ini}|$ for wash-in (blue solid line) and $|\mu_{\Delta_\tau}/\mu_{\rm bias}|$ for spontaneous (red dashed line) leptoflavorgenesis as functions of $T^{\rm dec}_{\ell \gamma} / T^{\rm dec}_{y_e}$ for the LFV process $\mu \to e \gamma$ (top panel) and $T^{\rm dec}_{\ell_L a} / T^{\rm dec}_{y_e}$ for the LFV process $\mu \to e a$ (bottom panel). The dark shaded region is excluded by current experimental bounds, while the light shaded region corresponds to the future expected sensitivity. 
    }
	\label{fig:b-l_aJBL}
\end{figure}

We numerically solve the transport equations with nonzero initial $\mu_e^{\rm ini}$ and obtain $q_{\Delta_\tau}$ for a given LFV coupling. The result is shown in Fig.~\ref{fig:b-l_aJBL}, which represents $\mu_{\Delta_\tau}/\mu_e^{\rm ini}$ as a function of the ratio between the decoupling temperatures of the LFV interaction and the electron Yukawa interaction, $T^{\rm dec}_{\ell \gamma} / T^{\rm dec}_{y_e}$, where $T^{\rm dec}_{y_e} \simeq 1.1 \times 10^5 \GeV$. The result is asymptotic to $\mu_{\Delta_\tau} = (79/213) \mu_e^{\rm ini}$ in the limit of large $T^{\rm dec}_{\ell \gamma} / T^{\rm dec}_{y_e}$. The dark shaded region is excluded because of the constraint on the corresponding $\mu^+ \to e^+ \gamma$ process. Future LFV searches will probe decoupling temperatures in the range represented by the light shaded region. If LFV should soon be discovered, the conversion factor from $\mu_{e}^{\rm ini}$ to $\mu_{\Delta_\tau}$ is of $\mathcal{O}(0.01)$. 

Setting $C^{ff'}_{La} = 1$, $C^{ff'}_{Ra} = 0$, the result based on the operator in Eq.~(\ref{eq:muea}) is shown in the lower panel of Fig.~\ref{fig:b-l_aJBL}. 
Although the form of transport equations is the same as the above case, the result is different because of the different temperature dependence of interaction rate $\gamma_{\ell_L a}$. 
The future experimental reach is shown by the light shaded region. 


\smallskip
\textit{\textbf{Spontaneous leptoflavorgenesis.}}\,---\,Finally, we consider another possibility to generate $\Delta_f$ 
by a mechanism similar to spontaneous baryogenesis~\cite{Cohen:1987vi,Cohen:1988kt} (see also Refs.~\cite{Chiba:2003vp,Takahashi:2003db,Domcke:2020kcp}). Now, we assume nonzero bias factors ($\mu_{\rm bias}^I \ne 0$) and vanishing initial charges ($\mu_C^{\rm ini} = 0$). 
For example, an axion coupling to an operator $\mathcal{O}^I$ of the form $-(a/f)\mathcal{O}^J$ results in $\mu_{\rm bias}^I = \dot{a}/f \delta_{IJ}$~\cite{Co:2020xlh,Domcke:2020kcp,Co:2020jtv}, while an axion coupling to a current $J_i$ of the form $(\dot{a}/f) J_i^0$ leads to $\mu_{\rm bias}^I = (\dot{a}/f) ({\bm n}^I)_i$. 
We assume that $\mu_{\rm bias}^I$ ($\dot{a}$) is nonzero until the LFV interactions are decoupled (see, \textit{e.g.}, Refs.~\cite{Co:2019wyp,Co:2019jts}).

From Eqs.~\eqref{eq:transport1-2} and \eqref{eq:transport2-2}, we see that $\mu_{\Delta_e} \neq 0$ is generated via the LFV interaction if $\mu_{\rm bias}^I$ is not perpendicular to $b_{\Delta_e}^I$ and $b_{\Delta_{\Delta_\tau}}^I$. We dub this scenario spontaneous leptoflavorgenesis. 
As an example, we consider the case with 
$\sum_I b^I_e \mu_{\rm bias}^I = -(1/2) \sum_I b^I_{\Delta_e} \mu_{\rm bias}^I =  \sum_I b^I_{\Delta_\mu} \mu_{\rm bias}^I =  \sum_I b^I_{\Delta_\tau} \mu_{\rm bias}^I  \equiv \mu_{\rm bias}$. 
If the LFV interactions are much stronger than the electron Yukawa interaction and decouple at a temperature higher than $T^{\rm dec}_{y_e}$, we obtain $\mu_{\Delta_\tau} = (481/639) \mu_{\rm bias}$ for both Eq.~(\ref{eq:egam}) and Eq.~(\ref{eq:muea}). 
This corresponds to the limit of large $T^{\rm dec}_{\ell \gamma/a} / T^{\rm dec}_{y_e}$. 
On the other hand,  if the LFV interactions decouple at a temperature lower than $T^{\rm dec}_{y_e}$, we obtain $\mu_{\Delta_\tau} =  (4/3) \mu_{\rm bias}$ for both Eq.~(\ref{eq:egam}) and Eq.~(\ref{eq:muea}). 
This corresponds to the limit of small $T^{\rm dec}_{\ell \gamma/a} / T^{\rm dec}_{y_e}$. In the intermediate regime,  we need to solve the transport equations numerically. 
The result is shown by the red dashed lines in Fig.~\ref{fig:b-l_aJBL}. Contrary to wash-in leptoflavorgenesis, we can obtain $\mathcal{O}(0.1\,\text{-}\,1)$ conversion factors even for small $T^{\rm dec}_{\ell \gamma/a} / T^{\rm dec}_{y_e}$.

Finally, we comment on the necessary condition for the spontaneous leptoflavorgenesis. 
We require a violation of lepton-flavor universality somewhere, such as in bias factors, equilibrated SM interactions, or CLFV interactions. 
Otherwise, the lepton flavor asymmetry cannot be produced.


\smallskip
\textit{\textbf{Discussion.}}\,---\,In this letter, we proposed wash-in and spontaneous leptoflavorgenesis.
In both cases, we need some UV physics to generate an asymmetry in the SM charges.
Though we are agnostic about the details of UV physics, the SM-charge generation through a phase rotation of a complex scalar condensate, \textit{i.e.,} the Affleck--Dine mechanism, is an interesting possibility. There, we generically expect the production of a relatively large asymmetry, which is suitable for leptoflavorgenesis as the conversion factor is suppressed by the charged-lepton Yukawa coupling.

We showed that almost any primordial charge can be reprocessed into a lepton flavor asymmetry if LFV is efficient in the early Universe, which in the end can lead to the present baryon asymmetry. Although we focused on operators inducing $\mu \to e \gamma$ and $\mu \to e a$, other processes are also interesting, as they will be extensively searched for in the near future, such as $\mu N \to e N$ by DeeMe~\cite{Teshima:2018ise}, COMET~\cite{COMET:2018auw}, Mu2e~\cite{Mu2e:2014fns}, PRISM~\cite{Kuno:2005mm}, and $\mu \to e e e$ by Mu3e~\cite{Blondel:2013ia}. Our results do not qualitatively change in these cases.

The anomalies and hints of lepton universality violation, \textit{viz.}\ muon $g-2$~\cite{Muong-2:2006rrc,Jegerlehner:2009ry,Muong-2:2021ojo} and $B$ meson decay~\cite{LHCb:2017avl,LHCb:2019hip,LHCb:2021trn}, may be explained by an operator similar to the LFV interactions with the flavor indices replaced by the identical flavors. One may generally think that LFV processes are linked by these anomalies~\cite{Feng:2001sq,Calibbi:2006nq,Giudice:2012ms,Glashow:2014iga}. In fact, much effort has been invested in order to construct UV models that explain the anomalies without introducing large LFVs~\cite{Greljo:2021xmg,Calibbi:2021qto,Ibe:2021cvf,Li:2021lnz,Buras:2021btx,BhupalDev:2021ipu,Greljo:2021npi,Hou:2021qmf}. 
We note that our mechanism works (even more efficiently) for suppressed but nonzero LFV interactions, including these models.

Finally, we comment on possible observable effects of the remnant lepton flavor asymmetry, which may be as large as $10^{-(4\cdots2)}$, depending on the scenario. The lepton flavor asymmetries remain until they are washed out by neutrino oscillations at a temperature of order $10 \MeV$~\cite{Wong:2002fa,Dolgov:2002ab}. In the literature, much attention has been paid to the observable effect of a total lepton asymmetry rather than lepton flavor asymmetries~\cite{Oldengott:2017tzj, Wygas:2018otj,Hajkarim:2019csy}. Recently, however, it has been pointed out that large flavor asymmetries may turn the QCD phase transition into a first-order phase transition~\cite{Gao:2021nwz}, which could lead to an observable signal in gravitational waves~\cite{Lerambert-Potin:2021ohy}. This provides another unique prediction of leptoflavorgenesis in addition to LFV interactions. 


\medskip\noindent\textit{Acknowledgments.}\,---\,KM was supported by MEXT Leading Initiative for Excellent Young Researchers Grant Number JPMXS0320200430. MY was supported by MEXT Leading Initiative for Excellent Young Researchers, and by JSPS KAKENHI Grant No.\ 20H0585, 20K22344, and 21K13910. This project has received funding from the European Union's Horizon 2020 Research and Innovation Programme under grant agreement number 796961, ``AxiBAU'' (KS).


\bibliographystyle{JHEP}
\bibliography{draft_1}


\clearpage
\appendix
\onecolumngrid


\renewcommand{\thesection}{S\arabic{section}}
\renewcommand{\theequation}{S\arabic{equation}}
\renewcommand{\thefigure}{S\arabic{figure}}
\renewcommand{\thetable}{S\arabic{table}}
\renewcommand{\thepage}{S\arabic{page}}
\setcounter{equation}{0}
\setcounter{figure}{0}
\setcounter{table}{0}
\setcounter{page}{1}


\begin{center}
\textbf{\Large Supplemental Material
}
\end{center}


In this Supplemental Material, we provide reduced transport equations by integrating out all SM spectator processes. Our conventions and notation in the following analysis will closely follow the discussion in~\cite{Domcke:2020quw} (see also Ref.~\cite{Domcke:2020kcp}).

Depending on the temperature of the SM plasma, some SM interactions are in equilibrium, while all other interactions are inefficient, thus leading to additional conserved quantities. We are interested in the transport equation around the decoupling temperature of the LFV interactions, which is of order $10^{4\cdots5} \GeV$. We can greatly simplify the transport equations in this regime, assuming that all interactions except for the electron Yukawa and LFV interactions are equilibrated. We denote the chemical potentials of the $16$ species in the Standard Model by $\left(\bm{\mu}\right)_i = \mu_i$, where the index runs through 
\begin{equation}
 i = e,\,\mu,\,\tau,\,\ell_e,\,\ell_\mu,\,\ell_\tau,\,u,\,c,\,t,\,d,\,s,\,b,\,Q_1,\,Q_2,\,Q_3,\,\phi\,.
\end{equation}
The multiplicity factor is given by 
\begin{equation}
\bm{g} = \qty( 1, 1, 1,  2, 2, 2,  3, 3, 3,  3, 3, 3,  6, 6, 6,  4 ) \,.
\end{equation}
We first derive the equilibrium values of these chemical potentials with all interactions in equilibrium, except for the electron Yukawa and LFV interactions. For this purpose, we need charge vectors ${\bm n}^I$ for the relevant interactions $I$, which can be found in, \textit{e.g.}, Refs.~\cite{Domcke:2020kcp,Domcke:2020quw}.
We adopt the convention of (S4$\,\text{-}\,$15) in Ref.~\cite{Domcke:2020quw}, such as 
${\bm n^{y_e}} = (-1,0,0,1,0,0,0,0,0,0,0,0,0,0,0,-1)$.
In addition, we use 
\begin{align}
\bm{n}^{\ell W/B, \mu e}    = & \left(-1,  0,  0, 0, 1, 0,  0,  0,  0,  0, 0,  0, 0, 0, 0, -1\right) \,, \\
\bm{n}^{\ell W/B, e \mu}    = & \left(0,  -1,  0, 1, 0, 0,  0,  0,  0,  0, 0,  0, 0, 0, 0, -1\right) \,, \\
\bm{n}^{\ell W/B, \tau e}    = & \left(-1,  0,  0, 0, 0, 1,  0,  0,  0,  0, 0,  0, 0, 0, 0, -1\right) \,, \\
\bm{n}^{\ell W/B, e \tau}    = & \left(0,  0,  -1, 1, 0, 0,  0,  0,  0,  0, 0,  0, 0, 0, 0, -1\right) \,, \\
\bm{n}^{\ell W/B, \tau \mu}    = & \left(0,  -1,  0, 0, 0, 1,  0,  0,  0,  0, 0,  0, 0, 0, 0, -1\right) \,, \\
\bm{n}^{\ell W/B, \mu \tau}    = & \left(0,  0,  -1, 0, 1, 0,  0,  0,  0,  0, 0,  0, 0, 0, 0, -1\right) \,, 
\label{eq:niu}
\end{align}
for the CLFV interactions of Eq.~(\ref{eq:egam}) 
and 
\begin{align}
\bm{n}^{\ell_R a, \mu e}    = & \left(-1,  1,  0, 0, 0, 0,  0,  0,  0,  0, 0,  0, 0, 0, 0, 0\right) \,, \\
\bm{n}^{\ell_L a, \mu e}    = & \left(0,  0,  0, -1, 1, 0,  0,  0,  0,  0, 0,  0, 0, 0, 0, 0\right) \,, \\
\bm{n}^{\ell_R a, e \tau}    = & \left(1,  0,  -1, 0, 0, 0,  0,  0,  0,  0, 0,  0, 0, 0, 0, 0\right) \,, \\
\bm{n}^{\ell_L a, e \tau}    = & \left(0,  0,  0, 1, 0, -1,  0,  0,  0,  0, 0,  0, 0, 0, 0, 0\right) \,, \\
\bm{n}^{\ell_R a, \tau \mu}    = & \left(0,  -1,  1, 0, 0, 0,  0,  0,  0,  0, 0,  0, 0, 0, 0, 0\right) \,, \\
\bm{n}^{\ell_L a, \tau \mu}    = & \left(0,  0,  0, 0, -1, 1,  0,  0,  0,  0, 0,  0, 0, 0, 0, 0\right) \,, 
\label{eq:niu}
\end{align}
for those of Eq.~(\ref{eq:muea}). 
Similarly, one can find vectors for conserved charges that are orthogonal to all charge vectors for the relevant interactions. For example, the hypercharge $Y$ and three lepton flavor charges $\Delta_f$ ($f = e, \mu, \tau$) are conserved in the SM. The electron charge (\textit{i.e.,} the right-handed electron number) is conserved if the electron Yukawa interaction is turned off. We denote the set of the linearly independent conserved charge vectors as ${\bm n}^C$ ($C = Y, \Delta_e, \Delta_\mu, \Delta_\tau, e$). If the initial chemical potentials of the SM fields are specified by a vector ${\bm \mu}^{\rm ini}$, the following quantities are constant,
\begin{equation}
 \left(\bm{n}^C \circ\bm{g}\right)
\cdot 
 {\bm \mu}^{\rm ini} =  \mu^{\rm ini}_C \quad ( = {\rm const.}), 
 \label{eq:conserved}
\end{equation}
for conserved charges $C$. 
Here, the $\circ$ symbol denotes the entrywise Hadamard product, such that $\left(\bm{n}^C \circ\bm{g}\right)_i = \left(\bm{n}^C \right)_i \left( \bm{g}\right)_i$.

Equipped with the above notation, 
we write the general transport equation as 
\begin{equation}
\label{generaltransportequation}
 \dot{\bm q} = - \frac{1}{T} \sum_I \gamma_I {\bm n}^I 
 \left(  {\bm n}^I \cdot{\bm \mu} - \mu_{\rm bias}^I \right), 
\end{equation}
where $\gamma_I$ is the linear response of the interaction $I$ or the rate per unit time-volume for an operator $\mathcal{O}^I$. 
We include a bias factor $\mu_{\rm bias}^I$, which is introduced in the spontaneous leptoflavorgenesis, as we explain in the main text. 
We rewrite the transport equations, assuming that all interactions except for the electron Yukawa and LFV interactions are equilibrated and that the latter two interactions are turned off. Together with the constrains of conserved charges, we obtain 
\begin{equation}
\label{eq:Mmu}
\mat{M}\,\bm{\mu}_{\rm eq} = \bm{m} \,,\qquad \mat{M} = \begin{pmatrix}\left(\bm{n}^I\right)^{\rm T} \\ \left(\bm{n}^C \circ\bm{g}\right)^{\rm T}\end{pmatrix} \,,\qquad \bm{m} = \begin{pmatrix} \mu_{\rm bias}^I \vspace{0.1cm}\\ \mu^{\rm ini}_C \end{pmatrix} \,,
\end{equation}
The first set of equations, associated with $\mu_{\rm bias}^I$, comes from the equilibrium conditions. The second set of equations, associated with $\mu^{\rm ini}_C$, comes from the constraints imposed by conserved charges with initial charges $\mu^{\rm ini}_C$, Eq.~(\ref{eq:conserved}). Here, we implicitly assume for simplicity that the set of vectors ${\bm n}^I$ are independent for all $I$. However, the charge vectors for the up and down Yukawa interactions are degenerate with that for the strong sphaleron and the former two interaction rates are comparable to each other. In such a case, some corrections arise with a combination of their interaction rates~\cite{Domcke:2020kcp}. In Eq.~\eqref{eq:Mmu}, we omit this for simplicity, while in the following analysis we take it into account (see Eq.~(\ref{eq:biase})). Equation~(\ref{eq:Mmu}) is solved by $\bm{\mu}_{\rm eq} = \mat{M}^{-1} \bm{m}$. Given this equilibrium solution, a certain charge $q_C$ specified by a vector ${\bm n^C}$ can be calculated from 
\begin{align}
 &q_C = \frac{T^3}{6} \frac{\mu_C}{T} 
= \frac{T^2}{6}  \left( {\bm n}^C \circ {\bm g} \right) \cdot {\bm \mu}_{\rm eq}. 
\label{charges}
\end{align}

Now we shall turn on the electron Yukawa as well as LFV interactions that violate some $\Delta_f$. For concreteness, we first consider the case based on the operators in Eq.~\eqref{eq:egam}. Since the decoupling temperature of the LFV interactions is comparable to that of the electron Yukawa interaction, we need to numerically solve the transport equations for $q_e$ and $\Delta_f$,
\begin{align}
 \dot{q}_e + 3 H q_e &= - \frac{1}{T} \gamma_e 
\left( \mu_e - \mu_{\ell_e} + \mu_\phi + \mu_{\rm bias}^e  
\right) 
- \sum_{f'} \frac{1}{T} \gamma_{\ell W/B}^{f'e} 
\left( \mu_{e} - \mu_{\ell_{f'}} + \mu_\phi + \mu_{\rm bias}^{\ell W/B, f'e} 
\right), 
\\
\dot{q}_{\Delta_f} + 3 H q_{\Delta_f} &= - \sum_{f'} \frac{1}{T} \gamma_{\ell W/B}^{ff'} 
\left( \mu_{f'} - \mu_{\ell_{f}} + \mu_\phi + \mu_{\rm bias}^{\ell W/B,ff'}
\right) - \sum_{f'} \frac{1}{T} \gamma_{\ell W/B}^{f' f} 
\left( - \mu_{f} + \mu_{\ell_{f'}} - \mu_\phi - \mu_{\rm bias}^{\ell W/B,f'f}
\right), 
\end{align}
where we take into account the Hubble expansion. Next, we need to rewrite $\mu_{\ell_f}$, $\mu_{f'}$, and $\mu_\phi$ in terms of $\mu_e$ and $\mu_{\Delta_f}$ making use of $\bm{\mu}_{\rm eq} = \mat{M}^{-1} \bm{m}$. Assuming a universal coupling $C_{\ell B}^{f f'} = C_{\ell B} \simeq 1.1 C_{\ell \gamma}$ and $C_{\ell W}^{f f'} = 0$ except for $f'=e$ for simplicity, we obtain 
\begin{align}
\label{eq:transport1}
 \dot{q}_e + 3 H q_e &= - \frac{1}{T} \gamma_e 
\left( \frac{711}{481} \mu_e + \frac{5}{13} \mu_{\Delta_e} 
- \frac{4}{37} \mu_{\Delta_\mu} - \frac{4}{37} \mu_{\Delta_\tau} 
- \sum_I b^I_e \mu_{\rm bias}^I 
\right), 
\\
\label{eq:transport2}
\dot{q}_{\Delta_e} + 3 H q_{\Delta_e} &= - \frac{1}{T} \gamma_{\ell B} 
\left( 
 \frac{474}{481} \mu_e 
+ \frac{12}{13} \mu_{\Delta_e} 
- \frac{15}{37} \mu_{\Delta_\mu} 
- \frac{15}{37} \mu_{\Delta_\tau} 
- \sum_I b^I_{\Delta_e} \mu_{\rm bias}^I 
\right), 
\\
\label{eq:transport3}
\dot{q}_{\Delta_\mu} + 3 H q_{\Delta_\mu} &= - \frac{1}{T} \gamma_{\ell B} 
\left( 
- \frac{237}{481} \mu_e 
- \frac{6}{13} \mu_{\Delta_e} 
+ \frac{115}{111} \mu_{\Delta_\mu} 
- \frac{70}{111} \mu_{\Delta_\tau} 
- \sum_I b^I_{\Delta_\mu} \mu_{\rm bias}^I 
\right), 
\\
\label{eq:transport4}
\dot{q}_{\Delta_\tau} + 3 H q_{\Delta_\tau} &= - \frac{1}{T} \gamma_{\ell B} 
\left( 
- \frac{237}{481} \mu_e 
- \frac{6}{13} \mu_{\Delta_e} 
- \frac{70}{111} \mu_{\Delta_\mu} 
+ \frac{115}{111} \mu_{\Delta_\tau} 
- \sum_I b^I_{\Delta_\tau} \mu_{\rm bias}^I 
\right).
\end{align}
We assumed $C_{\ell B}^{f e} = 0$ 
so that $q_e$ is not washed out by CLFV process for the wash-in leptoflavorgenesis. This assumption is not required for the spontaneous leptoflavorgenesis. 
Here, the bias factors are given by 
\begin{equation}
\label{eq:biase}
b_e^I = \bordermatrix{ & I = y_e   & y_\mu     & y_\tau    & y_u & y_c & y_t & y_d & y_s &  y_b & y_{s-d} & y_{b-s} & WS & SS  \cr
                         & -1 & \frac{-7}{481} & \frac{-7}{481} 
                         & \frac{-135}{481} r 
                         & \frac{-135}{481} r
                         & \frac{-135}{481} r
                         & \frac{-135}{481} (r-1)
                         & \frac{-135}{481} (r-1)
                         & \frac{-135}{481} (r-1)
                         & 0 & 0 
                         & \frac{66}{481} & \frac{-117+135r}{481}
                         \cr }\,,
\end{equation}
where $r \equiv \gamma_u / (\gamma_u + \gamma_d)$ comes from the correction from the up-quark Yukawa interaction.
We also have
\begin{align}
\label{eq:bias2}
 b_{\Delta_e}^I &= \frac{2}{3} \left( b_e^I +  \delta^I_{y_e} + \delta^I_{y_\mu} + \delta^I_{y_\tau} \right)  - \sum_{f'} \left( \delta^I_{\ell W/B, ef'} - \delta^I_{\ell W/B, f'e} \right), 
 \\
 b_{\Delta_\mu}^I &= \frac{1}{2} \left( -b_{\Delta_e}^I - \frac{4}{3} \delta^I_{ y_\mu} + \frac{4}{3} \delta^I_{ y_\tau} \right) - \sum_{f'} \left( \delta^I_{\ell W/B, \mu f'} - \delta^I_{\ell W/B, f'\mu} \right), 
 \\
\label{eq:bias4}
 b_{\Delta_\tau}^I &=  \frac{1}{2} \left( -b_{\Delta_e}^I + \frac{4}{3} \delta^I_{ y_\mu} - \frac{4}{3} \delta^I_{ y_\tau} \right) - \sum_{f'} \left( \delta^I_{\ell W/B, \tau f'} - \delta^I_{\ell W/B, f' \tau} \right), 
\end{align}
where $\delta^{I}_J$ is the Kronecker delta. The index $I$ runs for all standard model interactions in the summation of Eq.~(\ref{eq:transport1}) while it also includes the LFV interactions in Eqs.~(\ref{eq:transport2}\,\text{-}\,\ref{eq:transport4}) and Eqs.~(\ref{eq:bias2}\,\text{-}\,\ref{eq:bias4}).

For the case based on the operators in Eq.~\eqref{eq:muea}, the transport equations are reduced to 
\begin{align}
 \dot{q}_e + 3 H q_e &= - \frac{1}{T} \gamma_e 
\left( \frac{711}{481} \mu_e + \frac{5}{13} \mu_{\Delta_e} 
- \frac{4}{37} \mu_{\Delta_\mu} - \frac{4}{37} \mu_{\Delta_\tau} 
- \sum_I b^I_e \mu_{\rm bias}^I 
\right), 
\\
\dot{q}_{\Delta_e} + 3 H q_{\Delta_e} &= - \frac{1}{T} \gamma_{\ell_L a} 
\left( 
 \frac{474}{481} \mu_e 
+ \frac{12}{13} \mu_{\Delta_e} 
- \frac{15}{37} \mu_{\Delta_\mu} 
- \frac{15}{37} \mu_{\Delta_\tau} 
- \sum_I b^I_{\Delta_e} \mu_{\rm bias}^I 
\right), 
\\
\dot{q}_{\Delta_\mu} + 3 H q_{\Delta_\mu} &= - \frac{1}{T} \gamma_{\ell_L a} 
\left( 
 - \frac{237}{481} \mu_e 
- \frac{6}{13} \mu_{\Delta_e} 
+ \frac{26}{37} \mu_{\Delta_\mu} 
- \frac{11}{37} \mu_{\Delta_\tau} 
- \sum_I b^I_{\Delta_\mu} \mu_{\rm bias}^I 
\right), 
\\
\dot{q}_{\Delta_\tau} + 3 H q_{\Delta_\tau} &= - \frac{1}{T} \gamma_{\ell_L a} 
\left( 
- \frac{237}{481} \mu_e 
- \frac{6}{13} \mu_{\Delta_e} 
- \frac{11}{37} \mu_{\Delta_\mu} 
+ \frac{26}{37} \mu_{\Delta_\tau} 
- \sum_I b^I_{\Delta_\tau} \mu_{\rm bias}^I 
\right), 
\end{align}
where we assume a universal coupling $C_{La}^{ff'} =1$, $C_{Ra}^{ff'} = 0$ for simplicity. 
The bias factors are given by Eq.~\eqref{eq:biase} and 
\begin{align}
 b_{\Delta_e}^I &= \frac{1}{3} \left( 2 b_e^I + 2 \delta^I_{y_e} - \delta^I_{y_\mu} - \delta^I_{y_\tau} \right) - \sum_{f'} \left( \delta^I_{\ell_L a, ef'} - \delta^I_{\ell_L a, f'e} \right), 
 \\
 b_{\Delta_\mu}^I &= \frac{1}{2} \left( -b_{\Delta_e}^I + \delta^I_{ y_\mu} - \delta^I_{ y_\tau} \right) - \sum_{f'} \left( \delta^I_{\ell_L a, \mu f'} - \delta^I_{\ell_L a, f' \mu} \right), 
 \\
 b_{\Delta_\tau}^I &=  \frac{1}{2} \left( -b_{\Delta_e}^I - \delta^I_{ y_\mu} + \delta^I_{ y_\tau} \right) - \sum_{f'} \left( \delta^I_{\ell_L a, \tau f'} - \delta^I_{\ell_L a, f' \tau} \right). 
\end{align}
These relations imply that $q_{\Delta_f}$ vanishes if the electron Yukawa interaction as well as the LFV interactions are in equilibrium and if the bias factor respects the lepton flavor symmetries. 
This is not the case if the LFV interactions decouple while the electron Yukawa interaction is not efficient enough. 
In other words, we require a violation of lepton-flavor universality somewhere, such as in bias factors, SM interactions efficient at that temperature, or CLFV interactions. 
Otherwise, the lepton flavor asymmetry cannot be produced.

In general, a single process does not produce just one conserved quantity, partly because of spectator processes. This leads to an ambiguity in the definition of the decoupling temperature of an operator (or a freeze-out temperature of an asymmetry). A conventional definition of the decoupling temperature is $\gamma_C /H= 1$. However, we instead adopt the definition introduced in Ref.~\cite{Domcke:2020kcp}: $({\bm n}^I \circ {\bm g}^{-1}) \cdot {\bm n}^I \gamma_I/H = 1$. This gives $\gamma_e /H = (4/7) T^{\rm dec}_{y_e}/ T$ for the electron Yukawa interaction, 
$\gamma_{\ell B} / H = (4/7) (T/T^{\rm dec}_{\ell \gamma})^3$ for the interaction in Eq.~(\ref{eq:egam}), and 
$\gamma_{\ell_L a}/H = (T/T^{\rm dec}_{\ell a})$ 
for the interaction in Eq.~(\ref{eq:muea}) with $C_{La}^{ff'} = 1$, $C_{Ra}^{ff'} = 0$. 


\end{document}